\documentclass[aps,prl,superscriptaddress,reprint,nofootinbib]{revtex4-1}

\usepackage{graphicx,multirow}
\usepackage{bm}
\usepackage{amsmath}
\usepackage{amssymb}
\usepackage{amscd}
\usepackage{latexsym}
\usepackage{slashed}
\usepackage{color}
\usepackage{graphicx}
\usepackage{ulem}
\usepackage{color}
\usepackage[caption=false]{subfig}
\usepackage{endnotes}

\usepackage{lineno}

\begin{document}


\title{
Soft-gluon coupling and the TMD parton branching Sudakov form factor 
}

\author{A.~Bermudez Martinez}
\affiliation{Deutsches Elektronen-Synchrotron DESY, D 22607 Hamburg}
\author{F.~Hautmann}
\affiliation{Elementaire Deeltjes Fysica, Universiteit Antwerpen, B 2020 Antwerpen}
\affiliation{Theoretical Physics Department, University of Oxford, Oxford OX1 3PU}%
\author{L.~Keersmaekers}
\affiliation{Elementaire Deeltjes Fysica, Universiteit Antwerpen, B 2020 Antwerpen}
\author{A.~Lelek}
\affiliation{Elementaire Deeltjes Fysica, Universiteit Antwerpen, B 2020 Antwerpen}
\author{M.~Mendizabal Morentin}
\affiliation{Deutsches Elektronen-Synchrotron DESY, D 22607 Hamburg}
\author{S.~Taheri Monfared}
\affiliation{Deutsches Elektronen-Synchrotron DESY, D 22607 Hamburg}
\author{A.~M.~van Kampen}
\affiliation{Elementaire Deeltjes Fysica, Universiteit Antwerpen, B 2020 Antwerpen}

\begin{abstract}
\noindent
The evolution of transverse momentum dependent (TMD) distributions 
in Quantum Chromodynamics (QCD) 
can be formulated in a parton branching (PB) framework. 
We show that 
next-to-next-to-leading-logarithm (NNLL) accuracy can be achieved 
 in this framework  by using the concept 
 of  soft-gluon physical coupling.  
 We present results for the TMD distributions and 
 for the Collins-Soper kernel controlling rapidity evolution. 
 The results pave the way for  PB predictions at NNLL level 
 for physical observables at the Large Hadron Collider (LHC) 
 and future colliders.   
\end{abstract}




\maketitle

One of the essential elements of  
 the  high-precision  physics program at the 
Large Hadron Collider (LHC) and 
future high-energy experiments~\cite{Azzi:2019yne,LHeC:2020van,FCC:2018byv,Proceedings:2020eah,CEPCPhysicsStudyGroup:2022uwl} 
is a reliable,  accurate  
description of the initial state of the collision, involving  
hadron structure and 
initial-state QCD radiation. Depending on the kinematic 
regions probed, this can be treated 
by collinear~\cite{Kovarik:2019xvh} 
or transverse momentum dependent 
(TMD)~\cite{Angeles-Martinez:2015sea}
parton  distribution functions. 
The former provide an effective 1-dimensional picture of the  
hadronic initial states; the latter provide a 3-dimensional picture, 
which is needed in  phase-space regions near the 
kinematic boundaries such as the Sudakov   
region~\cite{Collins:1984kg} and high-energy 
region~\cite{Catani:1990eg}. 

The effort towards higher precision 
also affects the 
parton-shower Monte Carlo  generators~\cite{Buckley:2019kjt} employed for 
realistic event simulations at colliders.    
A large body of work is  devoted 
to improving  their logarithmic 
accuracy~\cite{Bewick:2019rbu,Nagy:2020rmk,Nagy:2020dvz,Forshaw:2020wrq,Holguin:2020joq,vanBeekveld:2024wws,vanBeekveld:2024qxs,FerrarioRavasio:2023kyg,vanBeekveld:2022zhl,Hamilton:2020rcu,Dasgupta:2020fwr,Herren:2022jej,Gellersen:2021eci} 
and exploring their systematic uncertainties associated with 
initial-state parton 
distributions~\cite{Nagy:2014oqa,Nagy:2020gjv,Dooling:2012uw,Hoche:2017hno,prestel:2020,vanBeekveld:2023chs,vanBeekveld:2022ukn,Mendizabal:2023mel,Frixione:2023ssx}.   
Refs.~\cite{Hautmann:2017xtx,Hautmann:2017fcj} 
propose the use  of TMD parton distributions in 
parton branching (PB) algorithms, 
providing a complementary approach to the 
TMD evolution~\cite{Angeles-Martinez:2015sea,Boussarie:2023izj}  
 that gives a method to incorporate TMD effects in parton showers and 
 Monte Carlo event generators.

The PB approach of Refs.~\cite{Hautmann:2017xtx,Hautmann:2017fcj} 
has so far been applied with leading-logarithm (LL) and 
next-to-leading-logarithm (NLL)  accuracy  in the  Sudakov region. 
For instance, it has been applied through  NLL 
  to the 
computation of $ Z / \gamma$ boson 
transverse momentum spectra~\cite{BermudezMartinez:2019anj,BermudezMartinez:2020tys} and  
 determination of TMD parton 
 distributions~\cite{Bubanja:2023nrd,Zhan:2024lym,Zhan:2025wup} from 
 Drell-Yan (DY) experimental data. 
 The NLL contributions 
 have been matched~\cite{BermudezMartinez:2019anj,Yang:2022qgk,Abdulhamid:2021xtt} 
 with next-to-leading-order (NLO)  perturbative corrections
 in the MCatNLO~\cite{Alwall:2014hca} framework. 
 The inclusion of logarithmic contributions to TMD evolution 
 from the high-energy (small-$x$) region 
 has also been studied~\cite{Hautmann:2022xuc,TaheriMonfared:2019bop}.
 However, PB contributions beyond NLL in the Sudakov region 
 have not been considered so far. 

 The purpose of this paper is to start the investigation of 
 the PB Sudakov form factor 
at  next-to-next-to-leading-logarithm (NNLL) accuracy and beyond. 
 To this end,  we use the 
 soft-gluon physical 
 coupling~\cite{Catani:2023kpw,Catani:2019rvy,Banfi:2018mcq} 
(which is the higher-order extension of  the 
Catani-Marchesini-Webber 
(CMW) result~\cite{Catani:1990rr})   
 in the Sudakov evolution of TMD parton distributions. 
  With  this approach, in this work we 
obtain the perturbative NNLL Sudakov coefficients in 
the PB TMD evolution  
and 
identify the non-perturbative Sudakov contributions  
at large distances.  

Besides improving the logarithmic accuracy of PB TMD evolution, 
this work allows us to 
observe features of the perturbative theory 
which appear for the first time at NNLL. 
Starting from NNLL,  the 
 double-logarithmic coefficient in the Sudakov form factor  
 is no longer proportional to the cusp anomalous 
dimension, due to the 
collinear anomaly~\cite{Becher:2010tm}. 
In other words, the NNLL accuracy cannot be achieved 
by the use of three-loop splitting functions but it can be achieved 
by the use of the physical soft-gluon coupling.
Working in the PB  TMD approach supplemented with the soft-gluon 
coupling, we relate 
the difference between the 
cusp and the Sudakov double-log coefficient
to the perturbative contribution to 
the Collins-Soper (CS) 
kernel~\cite{Collins:1984kg,Collins:1981va,Collins:1981uk}.   
We then perform the evaluation of 
the CS kernel at NNLL (including  perturbative and 
non-perturbative components), using the 
computational technique~\cite{BermudezMartinez:2022ctj}. 
 
 The results of this work will be applicable to studies of 
 collider observables. For the purpose of 
 such applications we recall that,  
 besides NLO matching  
 in the MCatNLO~\cite{Alwall:2014hca} framework,
 TMD multi-jet merging~\cite{BermudezMartinez:2021lxz,BermudezMartinez:2021zlg,BermudezMartinez:2022bpj},   
 in the M.L.~Mangano's (MLM)~\cite{Alwall:2007fs,Mangano:2006rw} method, 
 has also been developed;  
furthermore,  the PB TMD method is 
implemented in the open-source QCD  platform
  \verb+xFitter+~\cite{xFitter:2022zjb,Alekhin:2014irh}, while the 
  fitted TMD sets 
  are available in the {\sc TMDlib} library~\cite{Abdulov:2021ivr,Hautmann:2014kza} 
  and included in the  {\sc Cascade} Monte Carlo
event generator~\cite{CASCADE:2021bxe,CASCADE:2010clj}.   

In this paper we describe the main idea of this work and illustrate it with a few 
calculations at NNLL for the evolved TMD distributions and the CS kernel. 
Full details will be reported elsewhere~\cite{ALelekEtAll}. 
Results from this study have been presented earlier 
in Refs.~\cite{Martinez:2024twn,Lelek:2024kax}. 
In the following, we start by recalling basic features of the 
PB  Sudakov form factor; then we describe the soft-gluon coupling  
and discuss its role in the TMD evolution; finally we present 
NNLL numerical results.  

According to the 
PB method~\cite{Hautmann:2017xtx,Hautmann:2017fcj,BermudezMartinez:2018fsv}, 
TMD distributions fulfill evolution equations in terms of evolution kernels   
which can be expressed through the  Sudakov form factors  
\begin{eqnarray}
&& \Delta_a(\mu^2, \mu_0^2) = \exp\left( -\sum_b   \right. 
\nonumber\\ 
&& \left.   \times \int_{\mu_0^2}^{\mu^2}\frac{\textrm{d}\mu^{\prime 2}}{\mu^{\prime 2}} \int_0^{z_M} \textrm{d}z\; zP_{ba}^R(z,\alpha_s) \right)\;
\label{eq:RealSud}
\end{eqnarray}
where $z$ and $\mu^\prime$ are the branching variables, 
representing respectively the longitudinal momentum transfer 
and the mass scale at the branching; $z_M$ is the soft-gluon 
resolution scale~\cite{Hautmann:2017xtx}, 
characterizing resolvable and non-resolvable branchings; 
 $ P_{ba}^R(z,\alpha_s) $ are the real-emission 
splitting functions,  
computable in perturbation theory as power series expansions 
in the coupling   $\alpha_s$. The form factor 
  $  \Delta_a(\mu^2, \mu_0^2) $  may be interpreted as the probability for 
  parton $a$ to evolve from $\mu_0$  to $\mu$  without resolvable branchings. 
  By using unitarity and the momentum sum rule to relate 
  $ P_{ba}^R(z,\alpha_s) $ to the 
    virtual parts of the splitting functions~\cite{vanKampen:2021oxe},  
  and introducing the coefficients $k_a$ and $d_a$ of the 
  singular terms of the splitting functions for $z \to 1$,  
 the  form factor 
  $  \Delta_a(\mu^2, \mu_0^2) $ may be rewritten as
\begin{eqnarray} 
&& \Delta_a(\mu^2, \mu_0^2) = \exp\left( -\int_{\mu_0^2}^{\mu^2}\frac{\textrm{d}\mu^{\prime 2}}{\mu^{\prime 2}}  \right. 
\nonumber\\ 
&&  \left.   \times 
\left( \int_0^{z_M} 
\textrm{d}z \ k_a(\alpha_s) \frac{1}{1-z}   - d_a(\alpha_s)\right)\right)\;.
\label{eq:VirtSud}
\end{eqnarray}
The coefficients $k_a$ and $d_a$ are given as power 
series expansions in $\alpha_s$~\cite{Hautmann:2017fcj}, 
\begin{align}
\label{eq:k-and-d}
 k_a(\alpha_s)=\sum_{n=1}^{\infty} \left( \frac{\alpha_s}{2\pi}\right)^n k_a^{(n-1)},
  d_a(\alpha_s)=\sum_{n=1}^{\infty} \left( \frac{\alpha_s}{2\pi}\right)^n d_a^{(n-1)}\;.   
\end{align}
The one-loop contributions are 
\begin{align}
&k_q^{(0)} = 2 C_F,  &k_g^{(0)} = 2 C_A\;,  \\
&d_q^{(0)} =\frac{3}{2} C_F, &d_g^{(0)} =\frac{11}{6}C_A - \frac{2}{3}T_R N_f\;, 
\end{align}
where $C_A =3$, $C_F=4/3$, $T_R=1/2$, and $N_f$ is the number of flavors.   
The two-loop contributions are 
\begin{align}
k_a^{(1)} &= 2 C_a\left(C_A \left( \frac{67}{18}-\frac{\pi^2}{6}\right) - \frac{10}{9}T_RN_f\right), 
\label{k1a_cusp}
\\
d_q^{(1)} &= C_F^2 \left( \frac{3}{8}-\frac{\pi^2}{2}+6\zeta_3 \right) \nonumber\\
&+ C_FC_A \left( \frac{17}{24} + \frac{11}{18}\pi^2 - 3\zeta_3 \right) \nonumber\\
&- C_F T_R N_f \left( \frac{1}{6} + \frac{2}{9}\pi^2 \right), 
\label{d1q_b2msbar}
\\ 
d_g^{(1)} &= C_A^2 \left( \frac{8}{3} + 3 \zeta_3 \right) - \frac{4}{3} C_A T_R N_f - C_FT_R N_f\;,  
\label{d1g_b2msbar}
\end{align}
where $C_a=C_F$ for quarks and $C_a=C_A$ for gluons, and 
 $\zeta$ is the Riemann zeta function. 
The $k_a$ terms are double-logarithmic contributions, 
corresponding to the cusp anomalous dimension, 
while the $d_a$ terms are single-logarithmic. In the  
logarithmic power counting for the Sudakov evolution of TMD distributions, 
the LL accuracy is obtained by including the $k_a^{(0)}$ coefficients, 
and the NLL accuracy is obtained by including the 
$k_a^{(1)}$ and $ d_a^{(0)}$ coefficients~\cite{vanKampen:2021oxe}. 

We now proceed to make two 
transformations on the Sudakov form factor.  
First, to achieve NNLL accuracy   we appeal to the concept of 
soft-gluon physical coupling~\cite{Banfi:2018mcq,Catani:2019rvy}, 
which extends   the CMW result~\cite{Catani:1990rr} to higher order.  
To do this, we modify Eq.~(\ref{eq:VirtSud}) by the transformation 
$ \alpha_s \to    \alpha_s^{\text{phys}}$, 
with the  soft-gluon physical coupling given by 
\begin{equation} \label{eq:effectivecoupling}
    \alpha_s^{\text{phys}}=\alpha_s \left( 1 + \sum_{n=1}^\infty \mathcal{K}^{(n)} \left( \frac{\alpha_s}{2\pi}\right)^n  \right)\;, 
\end{equation}
where the first-order coefficient  is~\cite{Catani:1990rr}
\begin{align}
\label{K1_physcou}
\mathcal{K}^{(1)} = & C_A\left( \frac{67}{18}-\frac{\pi^2}{6} \right) -\frac{5}{9}N_f \; \  ,  
\end{align}
and the  second-order coefficient is~\cite{Banfi:2018mcq,Catani:2019rvy} 
\begin{align}
    \mathcal{K}^{(2)} = & C_A^2 \left( \frac{245}{24}-\frac{67}{9}\zeta_2 + \frac{11}{6}\zeta_3 + \frac{11}{5}\zeta_2^2 \right) \nonumber\\ 
    &+ C_FN_f \left( - \frac{55}{24}+2\zeta_3\right) 
    \nonumber \\
    &+ C_AN_f\left( - \frac{209}{108}+\frac{10}{9}\zeta_2 - \frac{7}{3}\zeta_3\right) 
    \nonumber\\ 
    & - \frac{1}{27}N_f^2 + \frac{\pi\beta_0}{2}\left( C_A\left( \frac{808}{27}-28\zeta_3 \right) - \frac{224}{54}N_f \right), 
\end{align}
where $\beta_0 = (11 C_A - 2 N_f)/(12 \pi)$. 
The replacement of the coupling constant by the 
physical soft-gluon coupling follows the analysis developed 
by~\cite{Banfi:2018mcq,Catani:2019rvy}, based on the 
resummation of soft multiple-parton radiation. It was first proposed at 
NLL in~\cite{Catani:1990rr} and later extended to the NNLL case.
We will see shortly that the role of the soft-gluon coupling in the PB TMD evolution 
is to guarantee that, provided two-loop splitting functions are used in the Sudakov 
form factor, NNLL accuracy is achieved, for both single-logarithmic and 
double-logarithmic terms.  

Second, as in Refs.~\cite{Bubanja:2023nrd,Hautmann:2019biw}  
for each branching evolution scale $\mu^{\prime }$ we 
classify 
parton emissions according to whether the emitted transverse momenta $q_\perp$ 
are above or below the semihard ``showering'' scale 
$q_0$~\cite{Webber:1986mc,Bassetto:1983mvz,Marchesini:1987cf} 
(of order ${\cal O}$ (1 GeV)).  By using  the kinematic relationship 
$q_{\bot}=(1-z)\mu^{\prime }$ from soft-gluon angular 
 ordering~\cite{Webber:1986mc,Dokshitzer:1987nm,Bassetto:1983mvz,Marchesini:1987cf,Catani:1990rr,Hautmann:2019biw}, 
  the integration over the branching's longitudinal momentum transfer $z$ in Eq.~(\ref{eq:VirtSud})  is split 
into two regions, 
\begin{eqnarray}
\label{regions-ab}
&& {\rm a})  : \;\;\;   z < z_{\rm{dyn}} = 1 - q_0 /  {\mu}^{\prime }  \  , 
\nonumber\\ 
&& {\rm b}) :  \;\;\;  z_{\rm{dyn}}  < z < z_M  \;  , 
\end{eqnarray} 
where $ z_{\rm{dyn}} = 1 - q_0 /  {\mu}^{\prime } $ is the 
dynamical resolution scale~\cite{Hautmann:2019biw,Hautmann:2025fkw} 
associated with the angular ordering.   
We refer to 
the contributions to evolution from regions a) and b), respectively, 
as the perturbative (P) and non-perturbative (NP) Sudakov components. 

We are now in a position to evaluate the form factor (\ref{eq:VirtSud}) including the 
soft-gluon physical coupling. For the perturbative Sudakov, 
we  apply to region a)  in Eq.~(\ref{regions-ab})  the  
mapping~\cite{Hautmann:2019biw}     
 induced by the angular-ordering relation 
   $q_{\bot}=(1-z)\mu^{\prime }$, which enables us  to 
    go from 
 the branching variables $(z , \mu^\prime) $ to $ ( q_\perp , z) $. Next we insert 
  two-loop splitting functions and physical coupling $\alpha_s^{\text{phys}}$. 
In order to avoid the double counting of the cusp contribution 
at order $\alpha_s^2$ between the coefficient $k_a^{(1)}$ in 
Eq.~(\ref{k1a_cusp}) 
and the coefficient $\mathcal{K}^{(1)}$ in Eq.~(\ref{K1_physcou}), 
we use the  subtracted form of the physical coupling 
$\alpha_s^{\text{subtr}} = 
\alpha_s^{\text{phys}} - \mathcal{K}^{(1)} \alpha_s^2 / ( 2\pi)$. 
  By performing the 
  longitudinal momentum integration, for the perturbative Sudakov form factor at NNLL we obtain 
\begin{eqnarray}
\label{eq-13} 
    \ln (\Delta_a^{(\text{P})}(\mu^2,q_0^2) ) &=&
    - 
    \int_{q_0^2}^{\mu^2} \frac{dq_\perp^2}{q_\perp^2} \frac{\alpha_s}{2\pi} \Bigl( 
{1 \over 2}   \ln \frac{\mu^2}{q_\perp^2}   
     k_a^{(0)} 
     \\      
   &+&     \frac{\alpha_s}{2\pi} {1 \over 2} k_a^{(1)}   \ln \frac{\mu^2}{q_\perp^2}  
-  d_a^{(0)}  
     \nonumber \\
    &+& \frac{\alpha_s^2}{(2\pi)^2}  \mathcal{K}^{(2)} k_a^{(0)} \frac{1}{2} \ln \frac{\mu^2}{q_\perp^2}   
 -  \frac{\alpha_s}{2\pi} d_a^{(1)}     
      +     ...\Bigr)\; . 
      \nonumber 
\end{eqnarray} 
The three lines in Eq.~(\ref{eq-13}) 
give, respectively,  the LL, NLL and NNLL contributions 
to the TMD parton branching Sudakov form factor. 
 In particular, the double-logarithmic coefficient of order 
 ${\cal O} (\alpha_s^3)$,  $\mathcal{K}^{(2)}  k_a^{(0)} \equiv A_a^{(3)}$,   
 is supplied by the soft-gluon coupling in the PB TMD evolution.   
The NNLL single-logarithmic coefficient,  
$-2d_a^{(1)} \equiv B_a^{(2
)}$,  
is  supplied, on the other hand,  
by the two-loop splitting functions  in 
the PB TMD evolution, with the explicit 
expressions given in 
Eqs.~(\ref{d1q_b2msbar}),(\ref{d1g_b2msbar})     
for quark and gluon channels.

\begin{figure}[pb]
\begin{center}
\includegraphics[width=7.7cm]{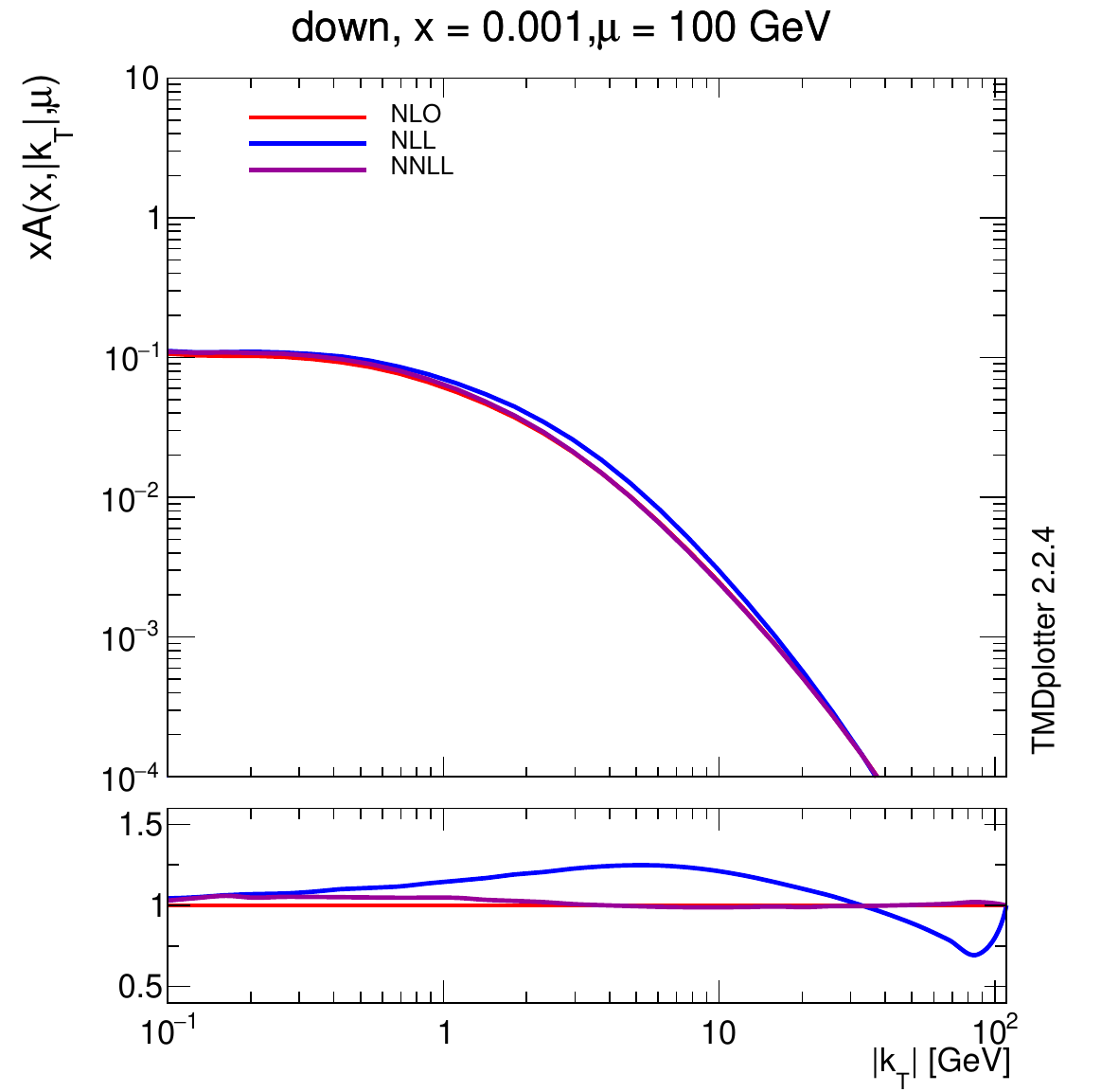}
\includegraphics[width=7.7cm]{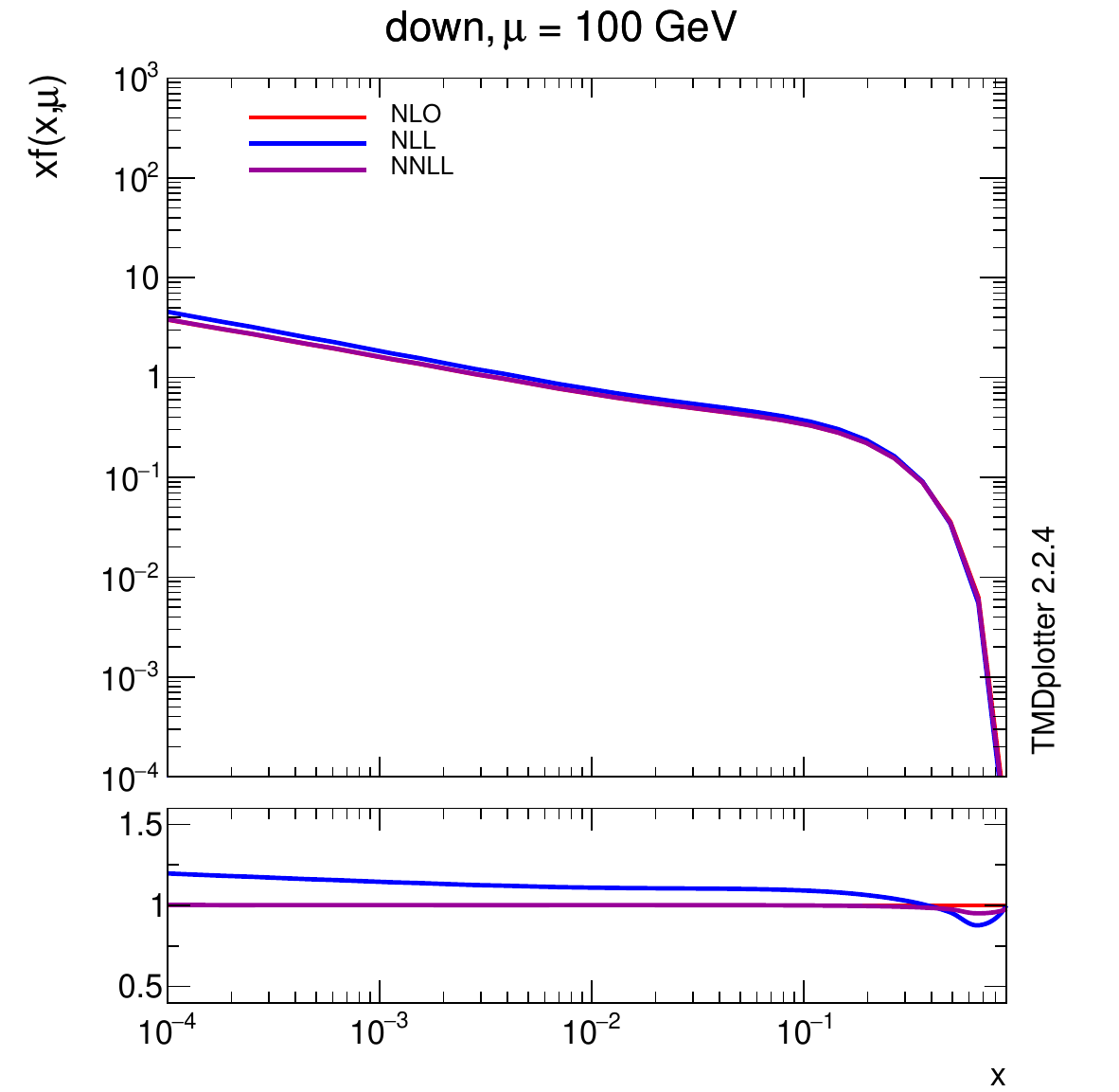}
\includegraphics[width=7.7cm]{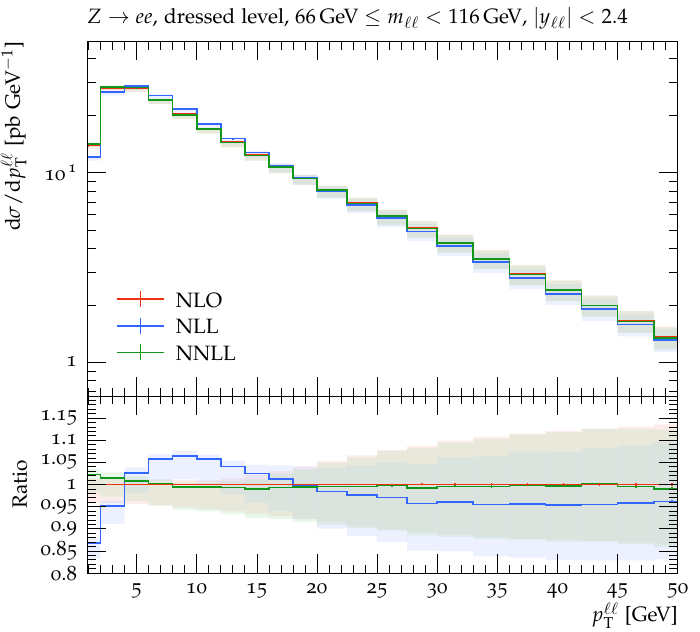}
\vspace*{8pt}
\caption{
 Impact of the soft-gluon physical coupling on down quark TMD (top), integrated TMD (middle) and DY $p_{\bot}$ spectrum at $\sqrt{s} = $ 8 TeV (bottom). We compare 
 NNLL and NLL results.} 
 \label{f1}
\end{center}
\end{figure}

We note that the PB method uses  the 
single-logarithmic, or $B$,  coefficients in the 
$\overline{MS}$ resummation scheme.  The 
general framework for resummation scheme 
transformations and the definition of commonly-used 
resummation schemes  is laid out in 
Refs.~\cite{Catani:2000vq,deFlorian:2001zd}. 
Here, changes in resummation schemes   
are expressed in terms of renormalization group 
transformations. In particular, the  
relationships between the 
$\overline{MS}$ resummation scheme 
used by PB 
and the commonly-used 
Drell-Yan (DY) and 
Higgs (H) schemes~\cite{Catani:2000vq,deFlorian:2001zd,Collins:2017oxh,Collins:2014jpa,Bizon:2017rah} 
can be obtained explicitly by using the results in 
Eqs.~(39),(40),(126) of Ref.~\cite{deFlorian:2001zd}. 
 These relationships are given by 
\begin{equation}
 B_q^{(2) \rm{DY}} -(-2)\cdot d_q^{(1)} = 16 C_F \pi \beta_0\left(\zeta_2-1\right)   \;
\end{equation}
and
\begin{equation}
 B_g^{(2) \rm{H}} -(-2)\cdot d_g^{(1)} = 16 C_A \pi \beta_0\left(\zeta_2+\frac{11}{24}\right)   \;.
\end{equation}
 
In Fig.~\ref{f1}  we present results of the numerical 
Monte Carlo implementation of the soft-gluon coupling 
in the PB TMD evolution. 
These are not intended as a detailed phenomenological 
study, but as an illustration that such studies will be 
feasible, based on the framework of this paper. 
The general set-up for the numerical evolution code is as 
in~\cite{BermudezMartinez:2018fsv}. In particular, 
 the strong coupling $\alpha_s$ 
is evaluated at the transverse momentum $q_\perp$, 
 according to angular 
 ordering~\cite{Bassetto:1983mvz,Catani:1990rr,Hautmann:2019biw}.  
The non-perturbative region b) of Eq.~(\ref{regions-ab})  
is treated by modeling the strong coupling as 
$\alpha_s = \alpha_s(\max(q^2_{\rm{cut}},{\bf q}_{\perp}^2))$, 
with $q_{\rm{cut}} $ on the order of the GeV. 
The resulting 
non-perturbative Sudakov form factor depends on this modeling of the 
low-$q_\bot^2$ region, as well as on the scale $q_0$ in Eq.~(\ref{regions-ab})  
and modeling of the soft-gluon resolution. 
The solid purple and blue curves give, 
respectively, the NNLL and NLL results.\footnote{We note that, 
while the  soft-gluon coupling  
coefficient $\mathcal{K}^{(1)}$  is subtracted in the NNLL 
calculation because it is proportional to  the cusp part $k_a^{(1)}$  of the 
two-loop splitting function,  the coefficient  $\mathcal{K}^{(1)}$ 
is used in the NLL calculation,  which employs one-loop 
splitting functions  together with  
$ \alpha_s^{\text{NLL}} = \alpha_s (1 + \mathcal{K}^{(1)} \alpha_s / (2 \pi))$. In  
the NLL calculation, the double-logarithmic Sudakov coefficient  
$A_a^{(2)}$ is obtained as 
$A_a^{(2)} \equiv  k_a^{(1)} = \mathcal{K}^{(1)} \cdot k_a^{(0)}   $.}

The top panel of 
Fig.~\ref{f1}  shows the transverse momentum dependence 
of the $d$-quark TMD distribution for 
given values of momentum fraction $x$ and evolution scale $\mu$.  
The middle panel shows the $x$ dependence of the same 
TMD distribution integrated over transverse momenta. The bottom 
panel shows predictions for the $Z$-boson transverse momentum distribution, 
obtained using NLO matching 
with MCatNLO as in~\cite{BermudezMartinez:2019anj,Yang:2022qgk}. 
For reference, we also show with the red curve the result of 
Ref.~\cite{BermudezMartinez:2019anj}. The uncertainty bands shown 
are those corresponding to variations of factorization and renormalization 
scales. We see that the primary effect of NNLL corrections is on 
 the shape of the $Z$-boson spectrum for $p_T$ below or around the 
peak, and that the region of higher $p_T$ above the peak is however also 
affected. 

The PB TMD method 
implies evolution in both mass and rapidity, in which the soft-gluon 
resolution scale plays the role of a rapidity 
cut-off~\cite{Boussarie:2023izj,Collins:1999dz,Collins:2000gd,Collins:2003fm,Hautmann:2007uw,Collins:2011zzd,Aybat:2011zv}. 
As a final application of our study, we next turn to the evaluation 
of the  CS kernel, i.e., the kernel for rapidity evolution.  The PB result for 
the rapidity-evolution kernel is presented next, and 
compared  with results from other methods.

\begin{figure}[pb]
\centerline{
\includegraphics[width=8cm]{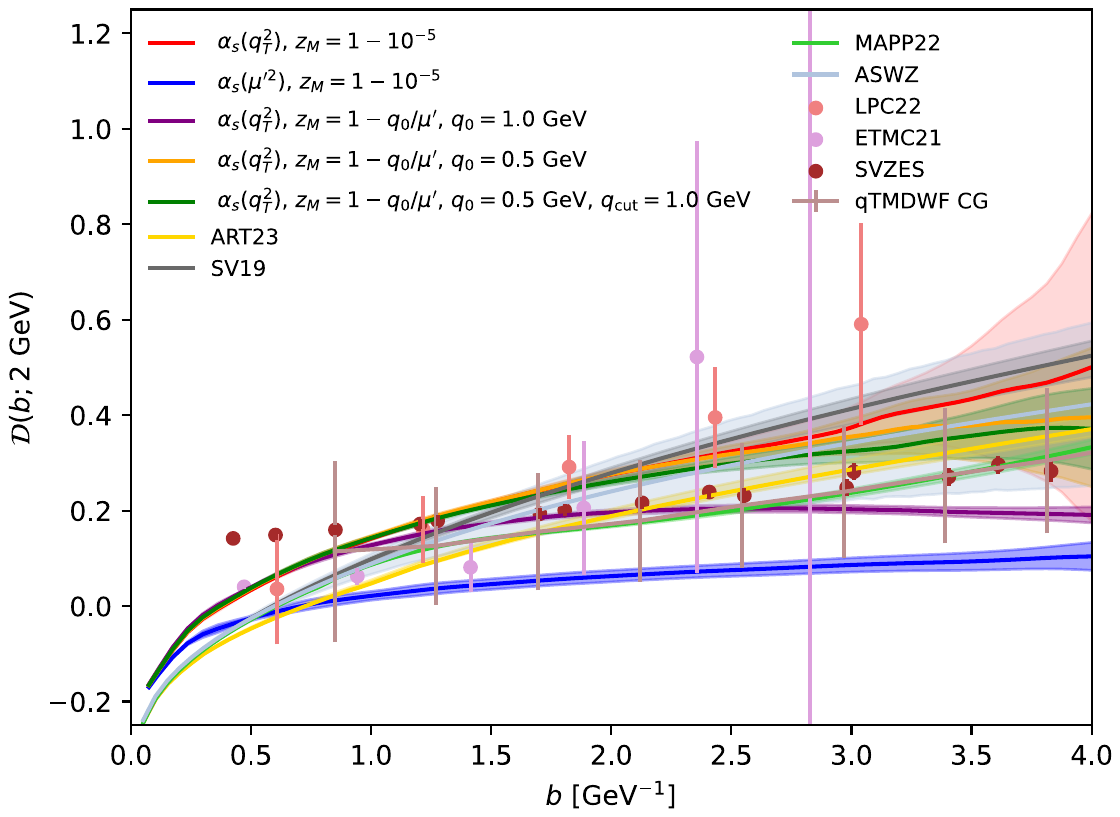}
}
\vspace*{8pt}
\caption{The $b$-dependence of the 
  CS kernel at $\mu = $ 2 GeV, obtained from NNLL predictions 
  with soft-gluon coupling (top five entries in the legend, 
  corresponding to different scenarios for resolution scale and 
  $\alpha_s$) and from several extractions 
  in the literature (next 
  eight entries in the legend, based on 
  fits to experimental data and on lattice calculations). 
 \label{f2}}
\end{figure}

The perturbative part of the 
CS kernel controls the relationship between 
the NNLL double-log coefficient,  obtained in the last line  
of Eq.~(\ref{eq-13}) 
from the PB TMD implementation of the soft-gluon coupling, and the 
 ${\cal O} (\alpha_s^3)$ $k$-coefficient in Eq.~(\ref{eq:k-and-d})~\cite{ALelekEtAll}.  
More precisely, by taking the derivative of the 
CS kernel with respect to transverse coordinate 
$b$~\cite{Collins:1981va,Collins:1981uk,Collins:2017oxh,Collins:2014jpa}, 
we have  
\begin{align} \label{eq:diffA3andk2}
    A_a^{(3)} - k_a^{(2)} &= C_a\pi\beta_0\left[ C_A\left(\frac{808}{27} - 28\zeta_3\right) - \frac{112}{27}N_f \right]  \; . 
\end{align}
The non-perturbative part of the CS kernel receives contribution, in our framework,  
from region b) of Eq.~(\ref{regions-ab}). It 
can be taken into account by using the 
technique~\cite{BermudezMartinez:2022ctj,BermudezMartinez:2023mrp}.  
This consists in extracting the CS kernel from ratios of appropriate Fourier transforms of 
physical  DY $p_T$ distributions computed at different masses.  
  We use this technique to study the CS kernel extracted from PB TMD 
  predictions including the soft-gluon physical coupling at NNLL. 

 In Fig.~\ref{f2} we show a summary plot, 
  reporting results for the CS kernel as a function of transverse coordinate 
  $b$ at evolution scale $\mu = $ 2 GeV.   
The plot in Fig.~\ref{f2}   presents 
  results obtained from our computations (corresponding to 
  different scenarios for the strong coupling and the 
  resolution scale, as indicated for  the five curves 
  at the top of the legend)    in comparison to several extractions from 
   the literature~\cite{Bacchetta:2022awv,Bacchetta:2024qre,Moos:2023yfa,Bury:2022czx,Scimemi:2019cmh,LPC:2022ibr,Schlemmer:2021aij,Li:2021wvl,Avkhadiev:2024mgd,Avkhadiev:2023poz,Bollweg:2024zet} 
   (based on fits to experimental data or on lattice calculations). 
   The uncertainty bands are computed using the 
   method proposed in~\cite{BermudezMartinez:2022ctj}. 
   The specific form 
of  $z_{dyn}$ in the resolution scale 
comes from the angular ordering of soft-gluon radiation. 
Thus, the corresponding results in Fig.~\ref{f2}  
directly probe the angular ordering picture. 

We observe 
that the curves with $\alpha_s ( q_\perp)$ (which  
fulfills angular ordering)  are close 
to one another  for $b < $ 1 GeV$^{-1}$ and start to differ for larger $b$ as an 
effect of the resolution scale, while the curve 
with $\alpha_s ( \mu)$ is already very different at small $b$. It is known 
from~\cite{BermudezMartinez:2019anj,Abdulhamid:2021xtt}, on the other 
hand, that $\alpha_s ( \mu)$ is already strongly disfavored by DY and jet 
data.  We also note the flattening behavior at large $b$ 
in the purple curve (dynamical resolution scale with $q_0 = 1$  GeV) compared to the 
red curve (fixed resolution scale). This is potentially of interest   
because, while traditionally fits to DY transverse 
momentum have assumed a quadratically rising 
large-$b$ behavior (see e.g.~\cite{Ladinsky:1993zn,Landry:1999an,Landry:2002ix,Konychev:2005iy,Bacchetta:2022awv}),  recent analyses 
have observed a preference for a 
flat large-$b$ behavior~\cite{Collins:2014jpa}, in a similar spirit to 
parton saturation in the $s$-channel 
picture~\cite {Hautmann:2007cx,Hautmann:2000pw} for 
partonic distribution functions 
  (see e.g.~\cite{Hautmann:2020cyp,Hautmann:2021ovt} 
in DY and~\cite{Boglione:2023duo} in $e^+ e^-$ fragmentation).

In conclusion, in this paper we study     
PB algorithms which  include the effects of TMD physics. 
We achieve perturbative NNLL accuracy 
by introducing the soft-gluon coupling in the 
TMD evolution and, taking into account 
non-perturbative Sudakov effects, we 
establish the relationship of the PB calculation 
at NNLL with the CS kernel throughout the range 
in transverse coordinate $b$, from short to long distances. 

This is the first computation of NNLL accuracy 
done with PB TMD techniques. It will impact future 
applications, given that PB TMD predictions have been 
successful in describing a wide range of collider processes, 
from DIS structure functions to DY spectra to multi-jets. Also, 
NNLL is the first logarithmic order sensitive to features 
of the perturbative theory such as the collinear anomaly. 
It thus provides a significant test of the method. 

The results obtained in this paper 
are relevant from the theoretical standpoint,  with 
the CS kernel being a major focus of 
non-perturbative QCD studies, e.g.~by lattice QCD methods, 
and from the phenomenological standpoint,  
as precision physics in electroweak boson production channels 
at the LHC (as well as future colliders) requires 
an accurate control of Sudakov and TMD dynamics 
in the low transverse momentum region.   

These results may be susceptible to being 
systematically extended to higher orders, as 
the notion of soft-gluon coupling also holds 
beyond order ${\cal O} (\alpha_s^3)$.  As 
regards applications to collider observables, 
we note that TMD matching and merging  
techniques are available (respectively 
with MCatNLO method and MLM method), so that 
predictions  become possible which include finite-order perturbative 
contributions together with the results of the present 
paper. 
As PB TMD algorithms can be implemented 
in parton-shower Monte Carlo event generators,  the 
results of this work can contribute to improve the 
accuracy of initial-state parton showers.

\vskip 0.3cm 

\noindent 
{\it Acknowledgments}. We thank 
A.~Banfi, T.~Becher, S.~Catani, H.~Jung and A.~Vladimirov 
for useful discussions. 
AL acknowledges funding by Research Foundation-Flanders (FWO) 
(application number: 1272421N).

\bibliography{bibliogSoftcou_rvsd3jun}



\end{document}